\documentclass[preprint,12pt]{elsarticle}
\usepackage{epsf,epsfig}
\usepackage[psamsfonts]{amssymb}
\usepackage{amsmath}
\usepackage{bm}
\usepackage{graphicx}
\usepackage{graphicx,epstopdf}
\usepackage{caption}
\usepackage{subcaption}
\usepackage{graphicx}

\newcommand{\be}{\begin{equation}}
\newcommand{\ee}{\end{equation}}
\newcommand{\bea}{\begin{eqnarray}}
\newcommand{\eea}{\end{eqnarray}}

\newcommand{\n}{\nonumber}

\begin{document}

\begin{frontmatter}
\title{Photon-added and photon-subtracted coherent states on a sphere}
\author{A Mahdifar $^{1,2}$}

\ead{mahdifar$_{-}$a@sci.sku.ac.ir}

\author{E Amooghorban $^1$}

\author{M Jafari $^1$}

\address{ $^1$ Department of Physics, Faculty of Basic Sciences, Shahrekord University, P.O. Box 115, Shahrekord 88186-34141, Iran.}
\address{$^2$ Department of Physics, Faculty of Science, University of Isfahan, Hezar Jerib, Isfahan, 81746-73441,
         Iran }

\begin{abstract}
In this paper, we construct the $m$-photon-added and $m$-photon-subtracted coherent states on a sphere.
These states are shown to satisfy the usual conditions of continuity
in the label, normalizability and the resolution of identity.
The preparation of the constructed states, as the states of radiation field
is considered.
We examine and analyze the nonclassical properties of these states, including the photon mean number, Mandel parameter and quadrature squeezing.
We find that these states are sub-Poissonian in nature, whereas the degree of squeezing is reduced (enhanced) by increasing $m$ for the photon-added (photon-subtracted) coherent states on a sphere.

\end{abstract}


\end{frontmatter}

\section{Introduction}

In the last decades, coherent states of the harmonic oscillator \cite{Ali 1} and generalized coherent states associated with various algebras \cite{Ali 2, Ali 3, Ali 4} have been playing an important role in various branches of physics. In particular, they have found
considerable applications in quantum optics and quantum information.

The coherent states, which defined as the eigenstates of the annihilation operator $\hat{a}$, are the quantum states with classical-like features. In this sense, they are the closest analogue to classical states. In contrast to the ordinary coherent states, the generalized coherent states which are known to exhibit some nonclassical properties,  have received an ever-increasing interest during the last decades.
Among the generalized coherent states, the so-called nonlinear
coherent states or $f$-deformed coherent states \cite{Ali 8} have attracted many interests in recent years
due to their applications in quantum optics and quantum technology. It is shown that these states
exhibit nonclassical properties, such as photon
antibunching \cite{Ali 5} or sub-Poissonian photon statistics \cite{Ali 6} and squeezing \cite{Ali 7}.
These states, which could be generated in the center-of-mass motion of an appropriately laser-driven
trapped ion \cite{Ali 10} and in a micromaser under intensity-dependent atom-field interaction \cite{Ali 11}, are associated with nonlinear algebras \cite{Ali 8, Ali 12}.

In Ref. \cite{Ali}, one of the authors of this paper (with Roknizadeh and Naderi) found that the deformed coherent states can be used to study a two dimensional harmonic oscillator on a sphere. In other words, a two-dimensional harmonic oscillator algebra can be treated as a deformed one-dimensional harmonic oscillator algebra. Moreover, the algebra
of an oscillator on a sphere represents a deformed version of the oscillator algebra in a flat space.
For the nonlinear coherent states on a sphere, which are special realizations of the deformed coherent states, one encounters with a finite-dimensional Hilbert space.
The influences of the curvature of the physical space on the algebraic structure and the physical properties of these coherent states is also analyzed in Ref \cite{Ali}. In addition, the modification of some nonclassical features of these states upon the propagation through the lossy media is examined in Refs. \cite{Asadi, IJPR}.

Recently, a proposal has been presented for the experimental realization of nonlinear coherent states on a sphere  in Ref. \cite{Ali Vo}. Based on the nonlinear
coherent states of the atomic center-of-mass motion of a trapped ion, one may control
the laser-ion interaction in such a way that nonlinear coherent states on a sphere are provided as motional
dark states of the system. Furthermore, the Rabi frequencies may be properly adjusted to simulate
and vary the curvature of the space somehow the nonlinear
coherent states under consideration are prepared.

On the other hand, a considerable attention is devoted to photon-added coherent states (PACSs) \cite{Sod 19, Sod 20, Sod 21, Sod 22}.
The PACSs  which first introduced
by Agarwal and Tara \cite{Sod 23}, represent interesting states generalizing both
the Fock states and coherent states. Indeed, they are obtained by repeatedly operating the photon creation
operator on an ordinary coherent state $|\alpha\rangle$.
The experimental realizations and applications of these states
were objects of extensive studies recently \cite{Six 8, Six 9}.
Unlike the operation of photon annihilation,
which maps a coherent state into another
coherent state, a photon excitation
of a coherent state changes it into
something entirely unlike the coherent state.
The PACSs, which can mathematically span a complete Hilbert space, reveal
nonclassical behavior different from that of the coherent states $|\alpha\rangle$.

%The various generalizations of the PACSs were also performed \cite{Sod 24, Sod 25}.

The generalizations of the PACSs have also received considerable attention in the quantum optics literature \cite{Hon 14, Hon 15, Hon 16}. For instance, recently Berrada \cite{Hon 17}, analogous to photon-added coherent states, introduced the photon-added spin coherent states. In addition, the photon addition and subtraction can be considered as one of the best approaches to generate and control the classicality in the quantum states \cite{Za sci}.

With the above background and regarding the fact that nonclassical lights play an important role in quantum information and communication, as the main purpose of this paper, we construct a special class of nonclassical states obtained by adding and subtracting photons to a nonlinear
coherent states on a sphere. We call these states the
photon-added coherent states (PACSs) and the photon-subtracted coherent states (PSCSs) on a sphere.
It will be shown that the aforementioned states can be derived by consecutive applying the deformed creation and annihilation operators on the nonlinear coherent states on a sphere. These states exhibit nonclassical properties, in the sense that their Mandel parameters become negative and quadrature amplitudes show squeezing.

The paper is organized as follows. In Sec. \ref{SCS}, we first briefly
review the previously introduced approach to mathematical
construction of the nonlinear coherent states on a sphere.
 In Sec. \ref{PACo}, the ordinary PACSs are introduced and then in Sec. \ref{SPACo}, the PACSs on a sphere are constructed in detail.
 The explicit form of the PSCSs on a sphere are given in Sec. \ref{SPSCo}.
 In Sec. \ref{Res}, the resolution of identity of
the introduced PACSs and PSCSs on a sphere are investigated.
 The possibility of the experimental realization of
PACSs and the PACSs on a sphere in a single-mode resonator is studied in
Sec. \ref{Gen}.
 We analyze numerically the nonclassical properties of the PACSs and the PACSs on a sphere
 and consider the influence of the number of added and subtracted photons and the curvature of the physical space on the mean number of photons, Mandel parameter and quadrature squeezing in Sec. \ref{Non}. Finally, a summary and some concluding remarks
are given in Sec. \ref{Sum}.

\section{Nonlinear coherent states on the surface of a sphere}\label{SCS}

Let us begin with the form of the nonlinear coherent states or $f$-deformed coherent states which are associated with an algebraic generalization of the coherent states. Based on this method, the
Weyl-Heisenberg creation and annihilation operators are replaced with their deformed ones as follows,
\bea\label{ty}
    \hat{A}
    &=&
    \hat{a} f (\hat{n}) =f ( \hat{n}+1) \hat{a},\n\\
    \hat{A}^{\dagger}
    &=&
    f^{\dagger} (\hat{n}) \hat{a}^{\dagger}= \hat{a}^{\dagger} f^{\dagger} (\hat{n}+1).
\eea
Here, $f(\hat{n})$ is a well-defined function of the number  operator $ \hat{n}=\hat{a}^{\dagger} \hat{a}$. Since usually the phase of $f$ is irrelevant, we can
choose $f$ to be real and nonnegative.
The operators $\hat{A}$ and $ \hat{A}^{\dagger}$ satisfy the following commutation relation:
 \bea
  [\hat{A},\hat{A}^{\dagger}]
  &=&
  (\hat{n}+1) f^{2}(\hat{n}+1)- \hat{n}f^{2}(\hat{n}).
 \eea
In \cite{Ali}, a two-dimensional oscillator on the surface of a sphere of radius  $R$ and curvature ${\lambda}$ $[\lambda =\frac{1}{R^{2}}]$ was studied and the relationship between deformation function in the theory of nonlinear coherent states and geometric structure of
physical space has been extracted.
By comparing the algebra of a two-dimensional oscillator on a sphere with a deformed oscillator algebra, it has been shown
that the two-dimensional oscillator can be considered as a one-dimensional  deformed harmonic oscillator with deformation function as follows:
 \begin{equation}
  f_{s} (n)=\sqrt{(N+1-n)} g(\lambda , n),
 \end{equation}
where
 \begin{equation}
  g(\lambda ,n) =\sqrt{\left(\lambda (N+1-n)+\sqrt{1+\frac{\lambda^{2}}{2}}\right)\left(\lambda n + \sqrt{1+\frac{\lambda ^{2}}{2}}\right)},
 \end{equation}
and $N$ is the dimension of finite-dimensional Fock space related to the harmonic oscillator algebra on the surface of a sphere. Due to the
finite dimensional Hilbert space of the oscillator on a sphere and making use of the formalism of constructing truncated coherent states \cite{7}, the
corresponding coherent states on a sphere can be obtained as follows:
 \begin{eqnarray}\label{scs}
  \vert \mu\rangle_{s} =N (\vert \mu \vert ^{2})^{-\frac{1}{2}} \exp (\mu \hat{A}^{\dagger}) \vert 0\rangle={N} (\vert \mu \vert
  ^{2})^{-\frac{1}{2}} \sum_{n=0}^{N} \sqrt {\left(\begin{array}{c}
   N\\
   n
  \end{array}\right)} [g (\lambda , n)]! \mu ^{n} \vert n\rangle,
 \end{eqnarray}
where $ N(\vert \mu\vert^{2})$ is the normalization coefficient expressed as follows:
 \begin{equation}
     N ( \vert \mu \vert ^{2}) = \sum_{n=0}^{N}{{\left(\begin{array}{c}
    N\\
    n
    \end{array}\right)}} \lbrace [g(\lambda , n)]!{\rbrace} ^{2}  ( \vert \mu \vert ^{2})^{n},
 \end{equation}
and by definition
       \begin{equation}\label{}
       [g(\lambda,0)]!=1,\hspace{0.5 cm}
       [g(\lambda,n)]!=g(\lambda,n)[g(\lambda,n-1)]!\hspace{0.1
       cm}.
       \end{equation}
It is evident that the coherent states $\vert \mu\rangle_{s} $ can be considered as a family of nonlinear coherent states corresponding to
curved space (sphere).

On the other hand, in the flat limit, $\lambda\rightarrow 0$, $g(\lambda,n)\rightarrow 1$ and the coherent states on a sphere
reduce to the coherent state on a flat space,
\begin{eqnarray}\label{fcs}
 |\mu\rangle_{f} &=&(1+|\mu|^{2})^{-N/2}\sum_{n=0}^{N}\sqrt{\left(\begin{array}{c}
N  \\
n
 \end{array}\right)}\hspace{.1 cm}\mu^{n}|n\rangle.
 \end{eqnarray}
It is worth noting that these coherent states are analogous to the spin coherent states that are
constructed by using the ${\frak{su}}(2)$ algebra \cite{Ali 26}.

 %%%%%%%%%%%%%%%\\\\\\\\\\\\\\\\\\\\\\\\\\\\\\\\\\\\\\\\\\\\\\\\\\\\\\\\\\\\\\\\\\\\\\\\\\\\\\\\\\\\\\\\\\\\\\\\\\\\\\%%%%%%%%%%%%%%%%%%%%%

\section{The photon-added coherent states}\label{PACo}

The PACSs were introduced theoretically in 1991 by Agarwal and Tara \cite{Sod 23}. These states are obtained by repeated application of
the photon creation operator on the coherent state as follows:
 \begin{equation}
    \vert \alpha , m\rangle ^{+} = \frac{(a^{\dagger})^{m}\vert \alpha \rangle} {\left(\langle \alpha \vert a^{m} (a^{\dagger})^{m}\vert
    \alpha \rangle \right)^{\frac{1}{2}}},
 \end{equation}
where $\vert \alpha \rangle$ is a standard coherent state and $m$ is the number of photon addition. It is clear that the PACS $\vert \alpha , m \rangle^{+} $, is reduced to the ordinary coherent state $\vert \alpha\rangle $ when $m$ approaches zero.
On the other hand, in the limit of $ \alpha \rightarrow 0 $, the state $ \vert \alpha , m\rangle$ becomes a number state $\vert
m\rangle$. Hence, the aforementioned state can be considered as an intermediate non-classical state which lies between the coherent
and the Fock states.

 \textit{Remark}. It should be noted that according to the algebraic definition of standard coherent states, $ \hat{a} \vert \alpha\rangle= \alpha\vert \alpha \rangle $, we can not construct a new state under the application of the annihilation operator $ \hat{a}$  on the coherent state $ \vert \alpha \rangle $. Therefore, on the contrary to PACSs, it is impossible to define the PSCSs based on the standard coherent states \cite{Zavatta}.
However, it is worth mentioning that since the dimension of the coherent states on a sphere (\ref{scs}) is finite, it is possible to define
not only the photon added but also the photon-subtracted coherent states on a sphere.

\section{The photon-added coherent states on a sphere}\label{SPACo}

In this section, we are intended to construct PACSs on a sphere.
The PACSs on a sphere can be defined as follows:
 \begin{equation}
    \vert \mu , m \rangle_{s} ^{+} = \frac{(A^{\dagger})^{m} \vert\mu \rangle_{s}} {\left(_{s}\langle \mu \vert A^{m} (A^{\dagger})^{m} \vert
    \mu \rangle_{s} \right)^{\frac{1}{2}}}.
 \end{equation}
Given the effect of the deformed creation operator on a sphere as follows
 \begin{equation}
    A^{\dagger}\vert n\rangle =g(\lambda,n+1) \chi_{n+1}^{N}\vert n+1\rangle,
 \end{equation}
where $ \chi_{n}^{N} =\sqrt{n(N+1-n)} $, the finite-dimensional PACSs are obtained as:
 \begin{equation}\label{PA}
    \vert \mu , m\rangle_{s}^{+}= C_{m}^{+}(|\mu|^{2})^{-\frac{1}{2}} \sum_{n=m}^{N} \sqrt{\left(\begin{array}{c}
    N\\
    n-m
    \end{array}\right)}\mu ^{(n-m)} g(\lambda ,n)!\  \left(\chi_{n}^{N}\right)_{m}^{+} \ \vert n\rangle.
 \end{equation}
Here, $  \left(\chi_{n}^{N}\right)_{m}^{+} $  is defined as:
 \begin{equation}
    \left(\chi_{n}^{N}\right)_{m}^{+}= (\chi_{n}^{N})(\chi_{n-1}^{N})\cdots(\chi_{n-m+1}^{N}),
 \end{equation}
and $C_{m}^{+} (|\mu|^{2})$  is the normalization coefficient derived by the following relation:
 \begin{equation}
    C_{m}^{+}(|\mu|^{2})=\sum_{n=m}^{N} {\left(\begin{array}{c}
    N\\
    n-m
    \end{array}\right)} \vert \mu \vert ^{2(n-m)} \left[ \left(\chi_{n}^{N}\right)_{m}^{+}\right]^{2} [g(\lambda , n)!]^{2}.
 \end{equation}
As it is seen from equation (\ref{PA}), the state $\vert \mu , m\rangle_{s}^{+}$ is a linear combination of
number states starting with $n= m$ and the first
$m$ number states
%$\left\{|n\rangle\right\}_{ 0\leq n< m} $,
are absent.
Therefore, the PACSs are belong to a $(N-m)$-dimensional Hilbert space, $\mathcal{H}^{+}_{(N-m)}$, defined as follows
 \be
       \mathcal{H}^{+}_{(N-m)}:=span\left\{|n \rangle\right\}_{n=m, \cdots ,N},
 \ee
which is the subspace of the $N$-dimensional Hilbert space $\mathcal{H}_{(N)}$ of the
coherent states on a sphere.
In other words, application of $(A^{\dagger})^{m}$, transfer the coherent states on a sphere from the total Hilbert space to its Hilbert subspace $\mathcal{H}^{+}_{(N-m)}$.

\section{The photon-subtracted coherent states on a sphere}\label{SPSCo}

In this section, we seek to construct PSCSs on a sphere. The effect of deformed annihilation operator on a sphere is given
by:
 \begin{equation}
    \hat{A}\vert n\rangle = g (\lambda , n )\ \chi_{n}^{N} \vert n-1\rangle.
 \end{equation}
Thus, the PSCSs on a sphere can be obtained as follows:
 \bea\label{PS}
    \vert \mu , m\rangle_{s}^{-}
    &\equiv&
    \frac{A^{m} \vert\mu \rangle_{s}} {\left(_{s}\langle \mu \vert (A^{\dagger})^{m} A^{m} \vert  \mu \rangle_{s} \right)^{\frac{1}{2}}}\\
    &=&
    C_{m}^{-}(|\mu|^{2})^{-\frac{1}{2}} \sum_{n=0}^{N-m} \sqrt{\left(\begin{array}{c}
    N\\
    n+m
    \end{array}\right)} \mu ^{(n+m)} g(\lambda , n+m)!  \left((\chi g)_{n}^{N}\right)_{m}^{-} \vert n \rangle.\n
 \eea
Here, $ \left((\chi g)_{n}^{N}\right)_{m}^{-}$  is defined as:
 \begin{equation}
    \left((\chi g)_{n}^{N}\right) _{m}^{-} =\left[\chi_{n+1}\ g(\lambda ,n+1)\right]\cdots\left[\chi_{n+m}\ g(\lambda , n+m)\right],
 \end{equation}
and the normalization coefficient, $ C_{m}^{-}(|\mu|^{2})$, is derived by the following relation:
 \begin{equation}
     C_{m}^{-}(|\mu|^{2}) =\sum_{n=0}^{N-m} {\left( \begin{array}{c}
    N\\
    n+m
    \end{array}\right)} \vert \mu \vert ^{2(n+m)} \left[g ( \lambda , n+m)!\right]^{2} \left[\left((\chi g)_{n}^{N}\right)_{m}^{-}
    \right]^{2}.
 \end{equation}
As it is seen from equation (\ref{PS}), the last
$m$ number states
%$\left\{|n\rangle\right\}_{ N-m < n\leq N} $,
are absent in the state $\vert \mu , m\rangle_{s}^{+}$.
Accordingly, by application of $A^{m}$, the coherent states on a sphere are shifted to $(N-m)$-dimensional Hilbert subspace $\mathcal{H}^{-}_{(N-m)}$ defined as follows
 \be
       \mathcal{H}^{-}_{(N-m)}:=span\left\{|n \rangle\right\}_{n=0, \cdots ,N-m}.
 \ee
\section{Resolution of identity}\label{Res}

According to the general definition of coherent states proposed by Klauder \cite{Kla 1963},
the set $|\alpha\rangle$ are called coherent states if
these states possess three minimal conditions: continuity
in the label $\alpha$, normalizability and the resolution of identity. The first
two conditions for the states $\vert \mu , m \rangle_{s}^{\pm}$ can be easily shown.

The resolution of identity is
defined in term of non-orthogonal projectors $|\alpha \rangle  \langle \alpha|$ as
\begin{equation}\label{35}
\int d^2\alpha |\alpha \rangle M(|\alpha|^2 ) \langle \alpha
|=\sum_{n}|n \rangle\langle n |=1,
\end{equation}
where $d^2\alpha=|\alpha|d|\alpha|d\theta$ and $M(|\alpha|^2 )$ is a proper measure which ensures the
completeness of these states.

For the coherent states on a flat space, equation (\ref{fcs}), we have
 \bea\label{36}
     \int d^2 \mu \vert \mu\rangle_{f}\
     M_{f}(|\alpha_{\mu}|^2 )\
     _{f}\langle \mu |&=&
      \pi \sum_{n=0}^{N} |n \rangle\langle n |\left(\begin{array}{c}
        N  \\
        n
        \end{array}\right)\n\\
        &\times&
        \int_{0}^{\infty} d (|\mu|^2 )\frac{ |\mu|^{2n}}{(1+| \mu|^2)^N }M_{f} (|\mu|^2 ).
 \eea
The completeness of the number states leads to the following expression:
\begin{equation}\label{38}
\int_{0}^{\infty} d (|\mu|^2 ) \frac{
|\mu|^{2n}}{(1+| \mu|^2)^N }M_{f}
(|\mu|^2 ) = \frac{1}{\pi\left(\begin{array}{c}
        N  \\
        n
        \end{array}\right) }.
\end{equation}
This allows us to choice a suitable measure function as
\cite{Ali 26}
\begin{equation}\label{39}
M_{f} (|\mu|^2 ) =\frac{N+1}{\pi}
\frac{1}{(1+|\mu|^2)^2}.
\end{equation}

In order to obtain the resolution of identify for the PACSs $\vert \mu , m\rangle_{s}^{+}$
and PSCSs $\vert \mu , m\rangle_{s}^{-}$, we first, respectively, define $(\lambda,m)$-deformed Binomial expansions as
 \bea\label{41}
     (1+x)_{\lambda,m}^{\pm} &=& \sum_{n=l^{\pm}}^{L^{\pm}} \left(\begin{array}{c}
        N  \\
        n
      \end{array}\right)_{\lambda,m}^{\pm} x^n,
 \eea
where
 \begin{eqnarray}
    \left\{ {\begin{array}{*{20}{c}}
    l^{+}=m &,& L^{+}=N  \\
    l^{-}=0 &,& \ \ \ \ \  L^{-}=N-m,
    \end{array}} \right.
 \end{eqnarray}
and
 \bea
 \left(\begin{array}{c}
        N  \\
        n
        \end{array}\right)_{\lambda,m}^{+} &=&\left(\begin{array}{c}
        N  \\
        n-m
        \end{array}\right) \left[\mu^{-m}\ g(\lambda ,n)!\  \left(\chi_{n}^{N}\right)_{m}^{+}\right]^{2},\n\\
         \left(\begin{array}{c}
        N  \\
        n
        \end{array}\right)_{\lambda,m}^{-} &=&\left(\begin{array}{c}
        N  \\
        n+m
        \end{array}\right)\left[\mu ^{m} g(\lambda , n+m)!  \left((\chi g)_{n}^{N}\right)_{m}^{-}\right]^{2}.
\eea
It is apparent that in the limiting case: $\lambda\rightarrow 0$ and $m=0$, these
deformed Binomial expansions reduce to the usual Binomial expansion.
By making use of these definitions, we can write
\begin{equation}\label{43}
   \vert \mu , m \rangle_{s}^{\pm} = \left[(1+|\mu|^{2})_{\lambda,m}^{\pm}\right]^{-\frac{N}{2}}\sum_{n=l^{\pm}}^{L^{\pm}}\sqrt{\left(\begin{array}{c}
N  \\
n
 \end{array}\right)_{\lambda,m}^{\pm}}\hspace{.1 cm}\mu^{n}|n\rangle.
 \end{equation}
Now, for the resolution of identity of the PASCs and PSCSs, we must have
\begin{equation}\label{44}
\int d^2 \mu|\mu,m \rangle_{s}^{\pm}\
M_{s}^{\pm}(|\mu|^2 )\ _{s}^{\pm}\langle \mu,m
|=1,
\end{equation}
or
  \begin{equation}\label{44}
        \pi \sum_{n=l^{\pm}}^{L^{\pm}} |n \rangle\langle n |\left(\begin{array}{c}
        N  \\
        n
        \end{array}\right)_{\lambda,m}^{\pm}\int_{0}^{\infty}d(|\mu|^2)
        \frac{ |\mu|^{2n}}{\left[(1+|\mu|^2)^{\pm}_{\lambda,m}\right]^{N}} M_{s}^{\pm}(|\mu|^2 )=1.
 \end{equation}
If we define the corresponding measure as
\begin{equation}\label{45}
  M_{s}^{\pm}(|\mu|^2 ) =\frac{N+1}{\pi}
\frac{1}{\left[(1+|\mu|^2)_{\lambda,m}^{\pm}\right]^2},
\end{equation}
and the deformed version of integral (\ref{38}) as
\begin{equation}\label{46}
\int_{0(\lambda^{\pm})}^{\infty} d (|\mu|^2 ) \frac{
|\mu|^{2n}}{\left[(1+|\mu|^2)_{\lambda,m}^{+}\right]^{N}} M_{s}^{\pm}
(|\mu|^2 ) = \frac{1}{\pi\left(\begin{array}{c}
        N  \\
        n
        \end{array}\right)_{\lambda,m}^{\pm} },
\end{equation}
the resolution of identity for the PACSs is obtained as
 \bea\label{47}
   \frac{N+1}{\pi}\int_{\lambda^{\pm}} \frac{ d^2\mu }{\left[(1+|\mu|^2)_{\lambda,m}^{\pm}\right]^2} |\mu,m\rangle_{s}^{\pm}
   \ _{s}^{\pm}\langle \mu,m|=
   \sum_{n=l^{\pm}}^{L^{\pm}} |n \rangle\langle n |=1_{\mathcal{H}^{\pm}_{(N-m)}}.
 \eea
Here, the second equality follows from the completeness of the number states in the Hilbert subspaces $\mathcal{H}^{+}_{(N-m)}$ and $\mathcal{H}^{-}_{(N-m)}$ of the Hilbert space $\mathcal{H}_{(N)}$ \cite{13 Popov}.
It is worth mentioning that the unity operator $1_{\mathcal{H}^{\pm}_{(N-m)}}$ is only required to be a bounded, positive operator with a densely defined inverse
\cite{33 Popov}.
In this manner, the over-completeness of PASCs and PSCSs have been derived in the Hilbert spaces $\mathcal{H}^{\pm}_{(N-m)}$.

%%%%%%%%%%%%%%%%%%%%%%%%%%%%%%%%%%%%%%%%%%%%%%%%%%%%%%%%%%%%%%%%%%%%%%%%%%%%%%%%%%%%%%%

\section{The photon-added and subtracted coherent states generation}\label{Gen}

In Ref. \cite{Vogel 2}, based on the transfer of atomic
coherence to the cavity field, a scheme for the generation of an arbitrary
field state in a single-mode resonator has been proposed. In this model, the atomic system
consists of a collection of two-level atoms which are initially prepared in a
linear superposition of the exited state $|e\rangle$ and the
ground state $|g\rangle$ as $|e\rangle+\varepsilon_k |g\rangle$.
Here, $\varepsilon_k$ denotes the complex amplitude coefficient for
the $k$th atom. These atoms coupled via the Jaynes-Cummings interaction to a resonant mode of the
electromagnetic field in a cavity which is initially in the
vacuum state. After the atom has passed through the
cavity, the measurement of the internal state of the atom leaves the quantum field unchanged in a pure state.
It is assumed
that after the passage of the $(k-1)$th atom and just before the
injection of the $k$th atom, the cavity field is in a state
 \be
   |\phi^{(k-1)}\rangle=\sum_{n}\phi_{n}^{(k-1)}|n\rangle.
 \ee
Immediately after leaving the atom, if finding it in the excited state, one then should
go back to the vacuum state and start the procedure again. On the other
hand, if the atom is detected in the ground state, we could then
continue the same process as before until more energy transfer from the atom to
the field and finally get the state of interest. After the
exit of the $k$th atom, the
new coefficients $\phi_{n}^{(k)}$ of the field state  are given in terms of the old coefficients $\phi_{n}^{(k-1)}$
as \cite{Vogel 2},
 \be
  \phi_{n}^{(k)}=\sin(g \tau_{k}\sqrt{n})\phi_{n-1}^{(k-1)}-\varepsilon_{k}\cos(g
  \tau_{k}\sqrt{n})\phi_{n}^{(k-1)},
 \ee
where $g$ is the atom-field coupling constant and $\tau_{k}$ is
the interaction time of the $k$th atom with the field.

Now, in a similar approach, we can prepare the finite-dimensional PACSs or PSCSs. To do so, we have to
find the combination of $N$ number states as follows
 \be
   |\phi^{(N-1)}\rangle=\sum_{n=0}^{N-1}\phi_{n}^{(N-1)}|n\rangle,
 \ee
which yields $\vert \mu , m\rangle_{s}^{\pm}$ after
the $N$th atom, which is prepared in an appropriate internal state
$|e\rangle+\varepsilon_N^{\pm} (\lambda, m) |g\rangle$,
has passed through the cavity and then has been detected in the ground state. Following
Ref. \cite{Vogel 2}, we construct a characteristic polynomial
equation for $\varepsilon_n^{\pm} (\lambda,m)$ of order $N$. By solving this equation
and choose one of $\varepsilon_N^{\pm} (\lambda,m)$ out of $N$ roots we can obtain a set of
$\phi_{n}^{(N-1)}$'s. In the next step, we take
$|\phi^{(N-1)}\rangle$ as a new desired state which must be
obtained by sending $N-1$ atoms through the cavity. For the state
$|\phi^{(N-1)}\rangle$, the similar calculations can be used (as for the state
$\vert \mu , m\rangle_{s}^{\pm}$) to obtain the parameter
$\varepsilon_{N-1}^{\pm} (\lambda,m)$ and state $|\phi^{(N-2)}\rangle$
with $N-1$ coefficients $\phi_{n}^{(N-2)}$.
Then, these instruction must be repeated until end up with the initial field state.
Eventually, a string of $\lambda$ and $m$ dependent complex numbers $\varepsilon_1^{\pm}(\lambda,m),
\varepsilon_2^{\pm}(\lambda,m), \cdots, \varepsilon_N^{\pm}(\lambda,m) $ defines
the internal states of a sequence of $N$ atoms which are supposed to inject
into the cavity. In this manner we can prepare the interested PACSs or PSCSs in a resonator.

It must be emphasized that detection of all the
$N$ atoms in the ground state $|g\rangle$ is a necessary condition to achieve
the PACSs or PSCSs into the cavity.
In fact, this conditional measurement is one of the
difficulties of the method, because the probability of finding
all $N$ atoms in the ground state after passing significantly reduces as $N$ becomes larger. However, by
using the degrees of freedom of the system, one can
optimize the probability of detecting the whole of $N$ atoms in the
ground state \cite{Vogel 2}.

\section{Quantum statistical properties of the photon-added and subtracted coherent states on a sphere}\label{Non}

After mathematical construction of the PACSs and PSCSs on a sphere, in the present section, we
investigate some of the quantum statistical properties of these states, including mean number of photons, Mandel parameter and quadrature squeezing.

\subsection{Photon-number distribution}

The photon distribution function and the mean number of photons are important tools to investigate the quantum statistical properties of quantum states. Based on this, in this section we examined these functions for the PACSs and PSCSs on a sphere.

\subsubsection{The photon-added}

By using Eq. (\ref{PA}), the probability of  finding  $n$ photons in the PACSs on a sphere can be obtained as:
 \begin{equation}\label{Pn+}
    (P_{n}) _{m}^{+} =  \frac{1}{C_{m}^{+}(|\mu|^{2})} {\left(\begin{array}{c}
    N\\
    n-m
    \end{array}\right)} \vert \mu \vert ^{2(n-m)}[g(\lambda , n)!]^{2} \left[(\chi_{n}^{N})_{m}^{+}\right]^{2}.
 \end{equation}

 \begin{figure*}[ht]
 \begin{minipage}[b]{0.5\linewidth}
 \centering
 \includegraphics[width=\textwidth]{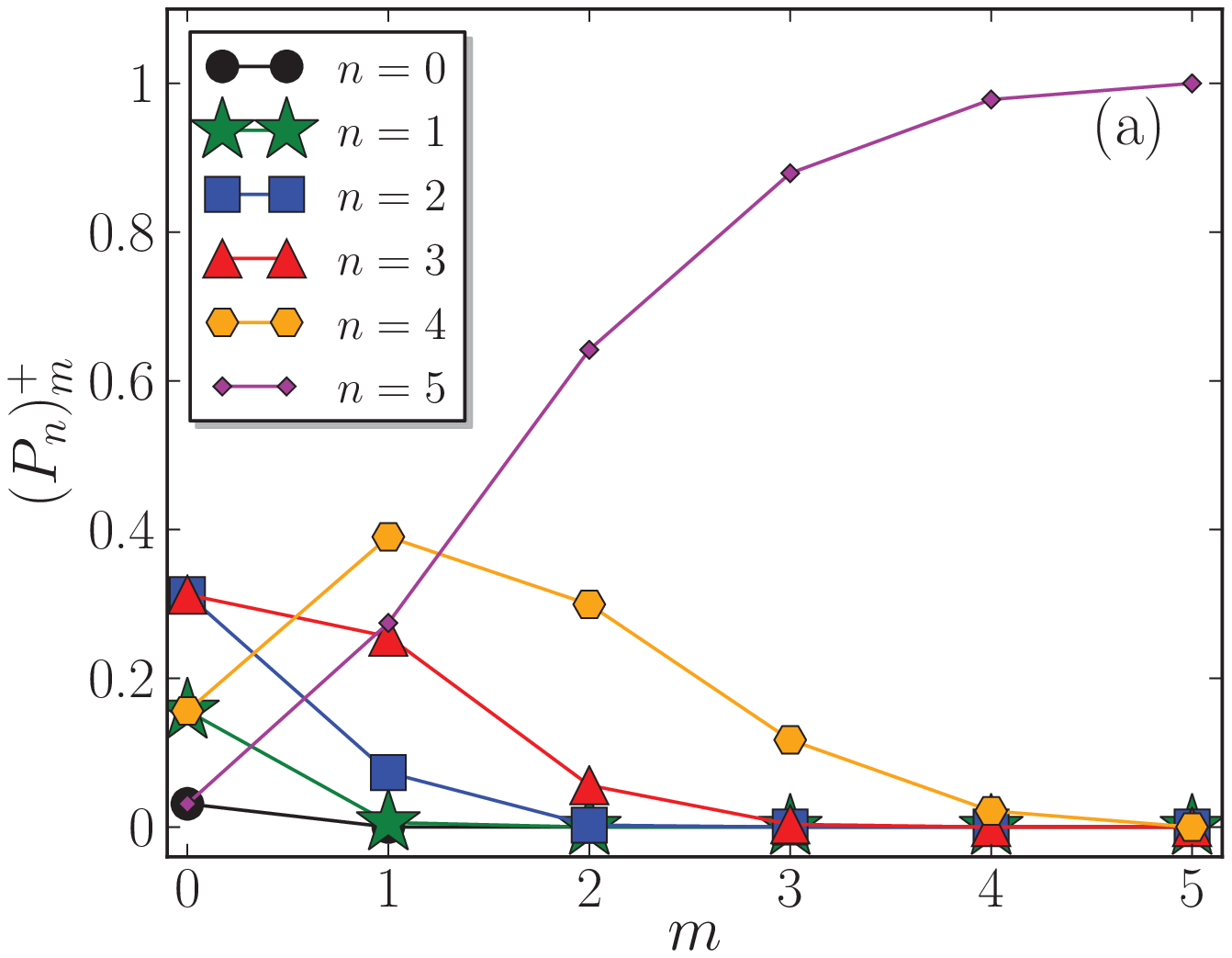}
 \end{minipage}
 \hspace{0cm}
 \begin{minipage}[b]{0.5\linewidth}
 \centering
 \includegraphics[width=\textwidth]{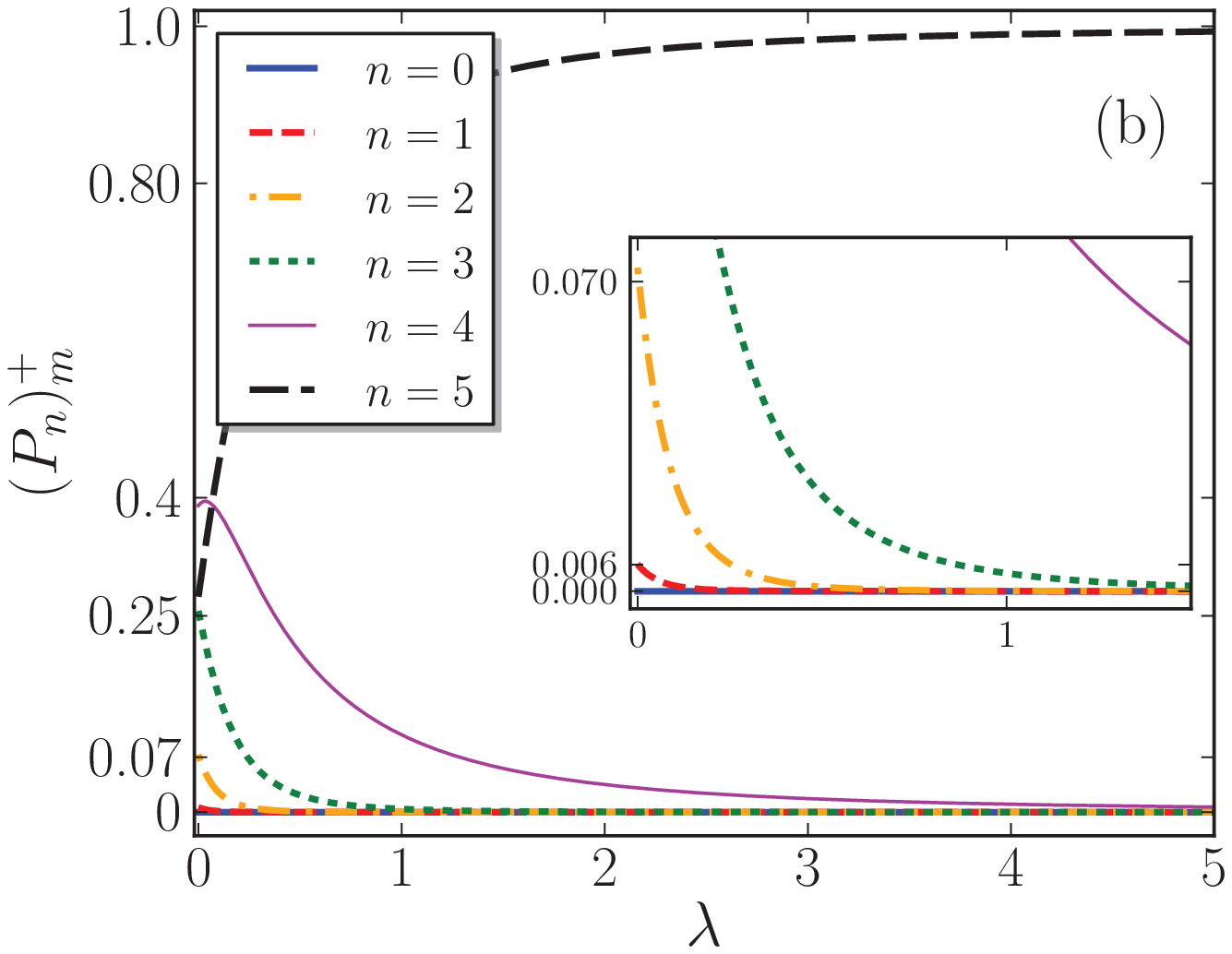}
 \end{minipage}
  \caption{(a) The probability distribution functions $\{(P_{n}) _{m}^{+}, n=0, 1, 2, 3, 4, 5\}$ of the PACSs on a sphere in terms of $m$ with $ N=5$, $\mu=1$ and
    $\lambda =0$.
    (b) The probability distribution functions of PACSs on a sphere versus $ \lambda $ for $N=5$, $\mu=1$ and $m =1$; $P_{0}$, thick-solid blue line; $P_{1}$, short-dashed red
  line; $P_{2}$, dash-dotted orange line; $P_{3}$, dotted green line; $P_{4}$, thin-solid violet line; $P_{5}$, long-dashed black line..}\label{11}
 \end{figure*}
The complexity of the above equation makes it difficult to predict the results analytically.
Therefore, in Fig. \ref{11}(a), the probabilities of finding $n$ photons in the PACSs are plotted in terms of $m$.
As  it is seen, the probabilities $(P_{n}) _{m}^{+}$ for $ n<m $ is zero.
This is as it should be, because the coefficient $(\chi_{n}^{N})_{m}^{+}$ become zero when $ n<m $.
It is also seen that by increasing $m$, the
PACS approaches to the $N$-quantum Fock state $ \vert N\rangle $. In Fig. \ref{11}(b), we show the probability $(P_{n}) _{m}^{+}$ as a function of $\lambda$, for $N=5$, $\mu=1$ and $m=1$.
In order to get some physical insight, let us consider
the limiting case: $\lambda\rightarrow \infty$. In this limit, the  probability distribution function of PACSs tends to $\delta_{n,N}$.
As a result, the PACS on a sphere approaches to the $N$-quantum Fock state $|N\rangle$, in the large curvature limit.

The mean number of photons in the state $\vert  \mu ,m\rangle_{s}^{+}$ is given by,
\begin{eqnarray}
\langle \hat{n} \rangle_{s}^{+} =_{s}^{+}\langle \mu , m\vert \hat{n} \vert  \mu ,m\rangle_{s}^{+}
                                =\sum_{n=m}^{N} n\ (P_{n})_{m}^{+}.
\end{eqnarray}
\begin{figure}[ht]
\centering
\includegraphics[scale=0.5]{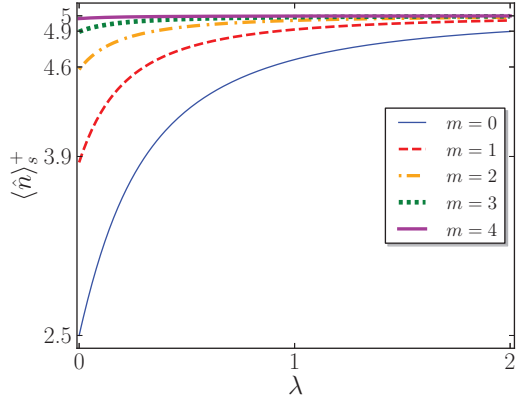}
\caption{Mean number of photons  for the PACSs on a sphere  versus $\lambda $ for $ N=5$ and $\mu=1$,
the thin-solid blue curve corresponds to $m=0$, the dashed red curve to $m=1$, the dot-dashed orange curve to $m=2$, the dotted green curve to $m=3$ and the thick-solid violet curve to $m=4$.}\label{33}
\end{figure}
In Fig. \ref{33}, the mean number of photons in the PACSs on a sphere is plotted in terms of $ \lambda $. As can be seen in this figure, the mean number of photons is increased by increasing the $ \lambda $  and in the limit of $\lambda\rightarrow\infty$ we have
\be
  \lim_{\lambda\rightarrow\infty}\langle \hat{n} \rangle_{s}^{+}\rightarrow N.
\ee
Besides, it is seen that by
increasing $m$, the mean number of photons increases more quickly.
These are in good agreement with the results presented in Figs. \ref{11}(a) and \ref{11}(b).\\

\subsubsection{The photon-subtracted}

In a similar manner with Eq. (\ref{Pn+}), by making use of the equation (\ref{PS}),
the probability of finding $n$ photons for the PSCSs on a sphere is given
by:
 \begin{equation}
    (P_{n})_{m}^{-}= \frac{1}{C_{m}^{-}(|\mu|^{2})} {\left( \begin{array}{c}
    N\\
    n+m
    \end{array}\right)} \vert \mu \vert ^{2(n+m)} \left[g(\lambda , n+m)!\right]^{2} \left[\left((\chi g)_{n}^{N}\right)_{m}^{-}\right]^2.
 \end{equation}

\begin{figure*}[ht]
 \begin{minipage}[b]{0.5\linewidth}
 \centering
 \includegraphics[width=\textwidth]{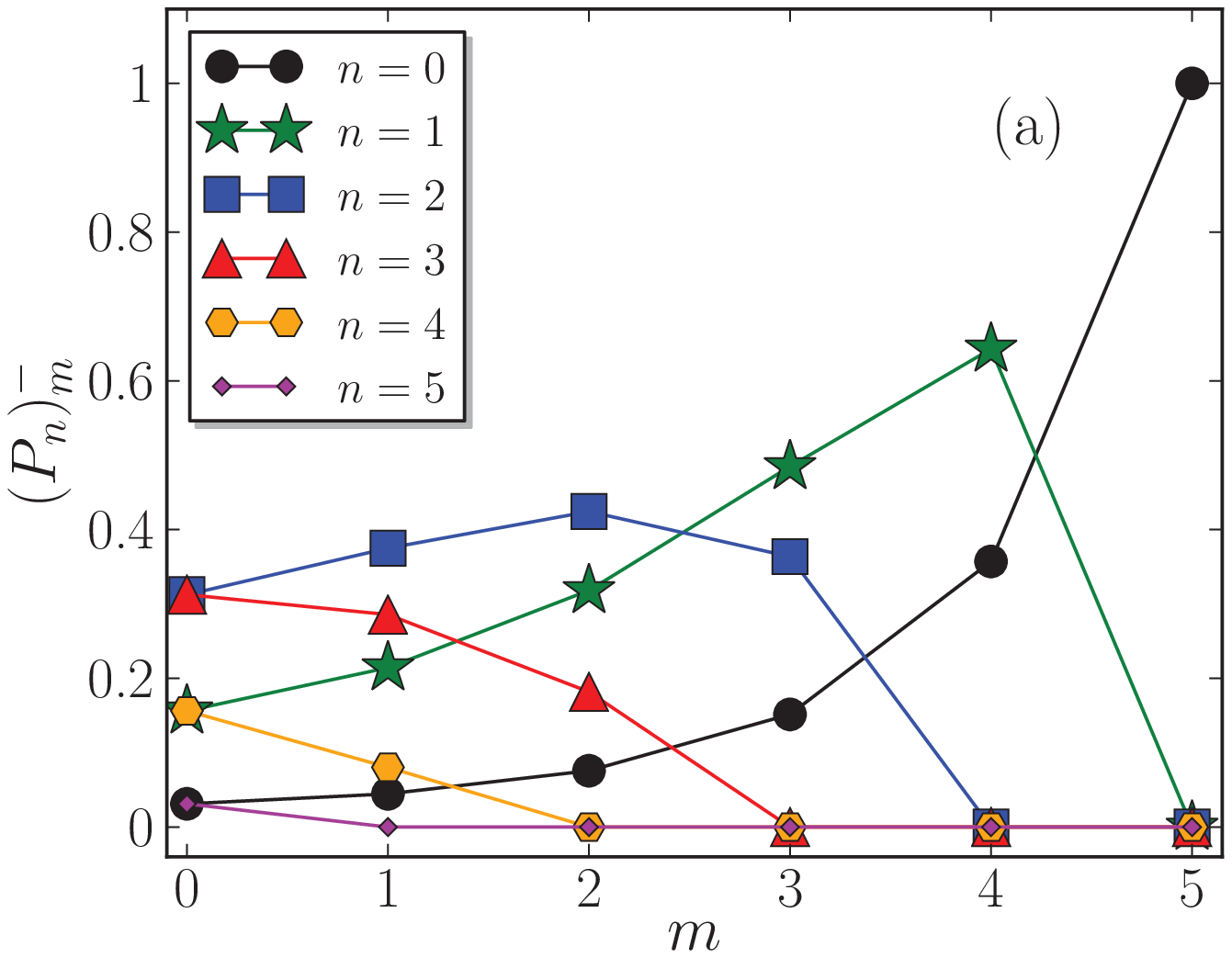}
 \end{minipage}
 \hspace{0cm}
 \begin{minipage}[b]{0.5\linewidth}
 \centering
 \includegraphics[width=\textwidth]{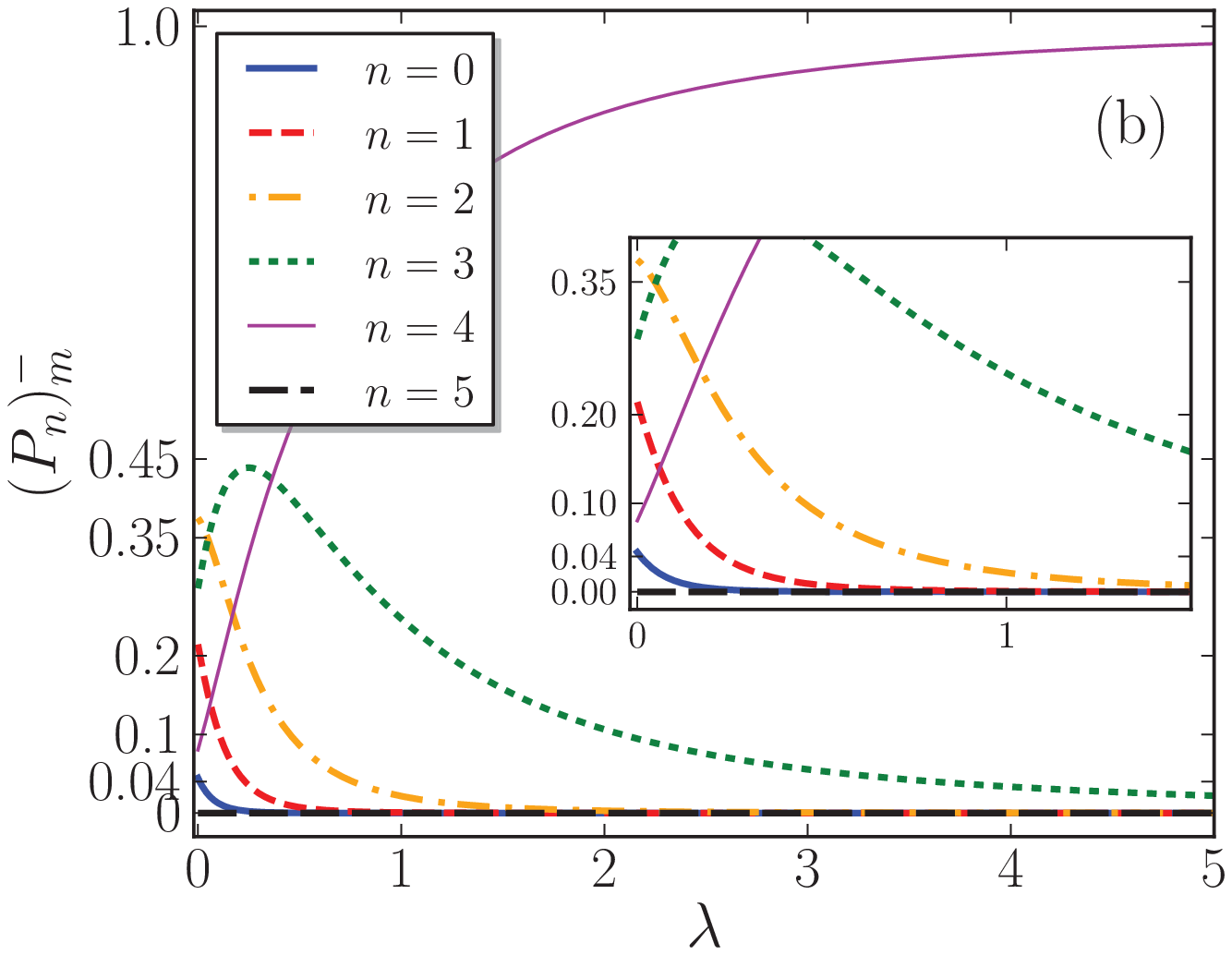}
 \end{minipage}
  \caption{(a) The probability distribution functions $\{(P_{n}) _{m}^{-}, n=0, 1, 2, 3, 4, 5\}$ of the PSCSs on a sphere in terms of $m$ with $ N=5$, $\mu=1$ and
    $\lambda =0$.
    (b) The probability distribution functions of PSCSs on a sphere versus $ \lambda $ for $N=5$, $\mu=1$ and $m =1$; $P_{0}$, thick-solid blue line; $P_{1}$, short-dashed red
  line; $P_{2}$, dash-dotted orange line; $P_{3}$, dotted green line; $P_{4}$, thin-solid violet line; $P_{5}$, long-dashed black line..}\label{44}
 \end{figure*}

%\begin{figure}[ht]
%\centering
%\includegraphics[scale=0.5]{Pn-m}
%\caption{The probability distribution functions $\{(P_{n}) _{m}^{-}, n=0, 1, 2, 3, 4, 5\}$ of PSCSs on a sphere in %terms of $m$ for $N=5$, $\mu=1$ and $\lambda=0
%$.}\label{44}
%\end{figure}

In Fig. \ref{44}(a), we show the probabilities of finding $n$ quanta in the state $\vert \mu ,m \rangle^{-} $, in terms of $m$.
As is seen, the probability  $(P_{n}) _{m}^{-}$ for $ n > (N-m) $ is zero and by increasing $m$, the
PSCS eventually approaches to the $0$-quantum Fock state $ \vert 0\rangle $,
because for $m_{max}=N=5$ we have $(P_{0})_{5}^{-}=1$, while $(P_{n})_{5}^{-}$ is zero for other values of $n$.
In Fig. \ref{44}(b), we show $(P_{n}) _{m}^{-}$ as a function of $\lambda$, for $N=5$, $\mu=1$ and $m=1$.
As it is seen, in the limit of $\lambda\rightarrow \infty$, the  probability distribution function of PSCSs tends to $\delta_{n,N-m}$.
As a result, by increasing $\lambda$, the PSCSs on a sphere approaches to the state $|N-m\rangle$.

The mean number of photons in the PSCS on a sphere is given by:
 \begin{eqnarray}
    \langle \hat{n} \rangle_{s}^{-}=_{s}^{-}\langle \mu , m\vert \hat{n} \vert  \mu ,m\rangle_{s}^{-}
                                   =\sum_{n=m}^{N} n\ (P_{n}) _{m}^{-}.
 \end{eqnarray}

\begin{figure}[ht]
\centering
\includegraphics[scale=0.5]{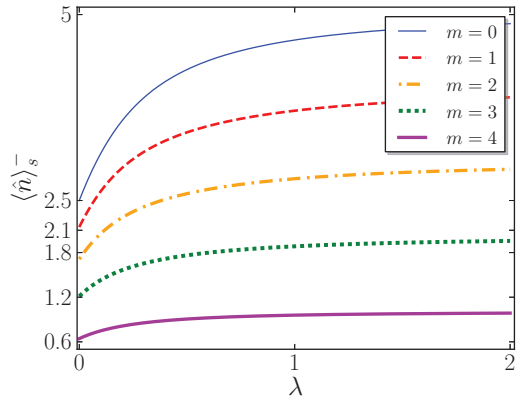}
\caption{Mean number of photons in the PSCSs on a sphere versus $\lambda$ with $N=5$ and $\mu=1$,
 the thin-solid blue curve corresponds to $m=0$, the dashed red curve to $m=1$, the dot-dashed orange curve to $m=2$, the dotted green curve to $m=3$ and the thick-solid violet curve to $m=4$.}\label{55}
\end{figure}

In Fig. \ref{55}, the mean number of photons $\langle \hat{n} \rangle_{s}^{-}$ is plotted in terms of $ \lambda $. The results show that all curves follow the same trend. The
mean number of photons first increases with the increase of $ \lambda $ and then tends to a constant value. Additionally, it can be found that by increasing $m$, the mean number
of photons is decreasing.
It is worth noting that these results are consistent with the results in Figs. \ref{44}(a) and \ref{44}(b).

\subsection{Mandel parameter}

Mandel parameter is frequently used to measure the deviation of a distribution from the Poissonian distribution and defined as
 \begin{equation}
    Q=\frac{(\triangle n)^{2} -\langle \hat{n} \rangle}{\langle \hat{n} \rangle}.
 \end{equation}
For the conventional coherent state with the Poissonian distribution, the variance of the number operator is
equal to its mean value and consequently $ Q=0 $, while the Mandel parameter is positive for a super-Poissonian distribution (photon bunching) and is negative for a
sub-Poissonian distribution (photon antibunching).\\
In this section, the Mandel parameter is examined for the PACSs and the PSCSs on a sphere. However, due to the complexity of the final form of this equation, we do not attempt to obtain its analytic form. Instead, we numerically study the Mandel parameters for these states.

\begin{figure*}[ht]
\begin{minipage}[b]{0.5\linewidth}
\centering
\includegraphics[width=\textwidth]{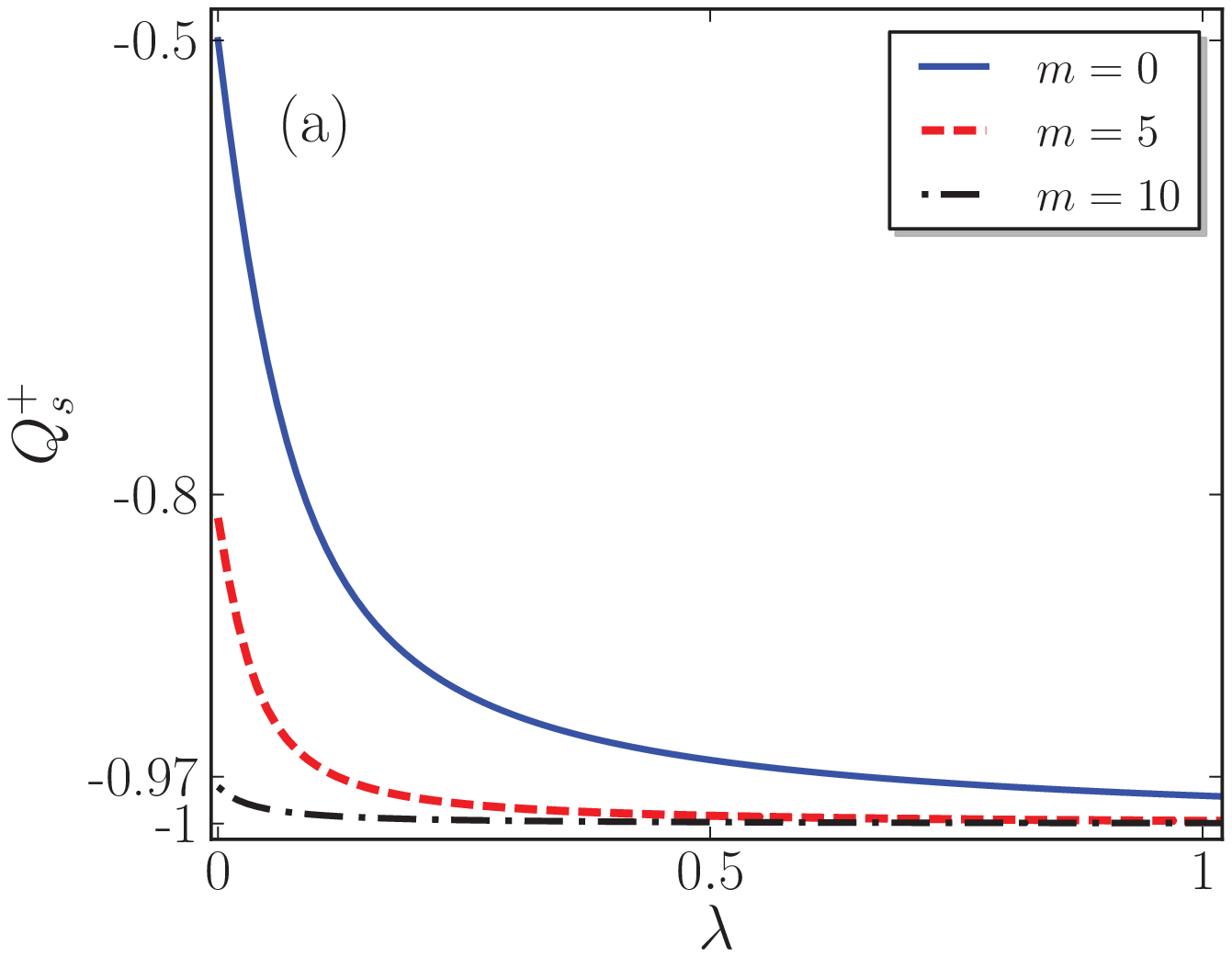}
\end{minipage}
\hspace{0cm}
\begin{minipage}[b]{0.5\linewidth}
\centering
\includegraphics[width=\textwidth]{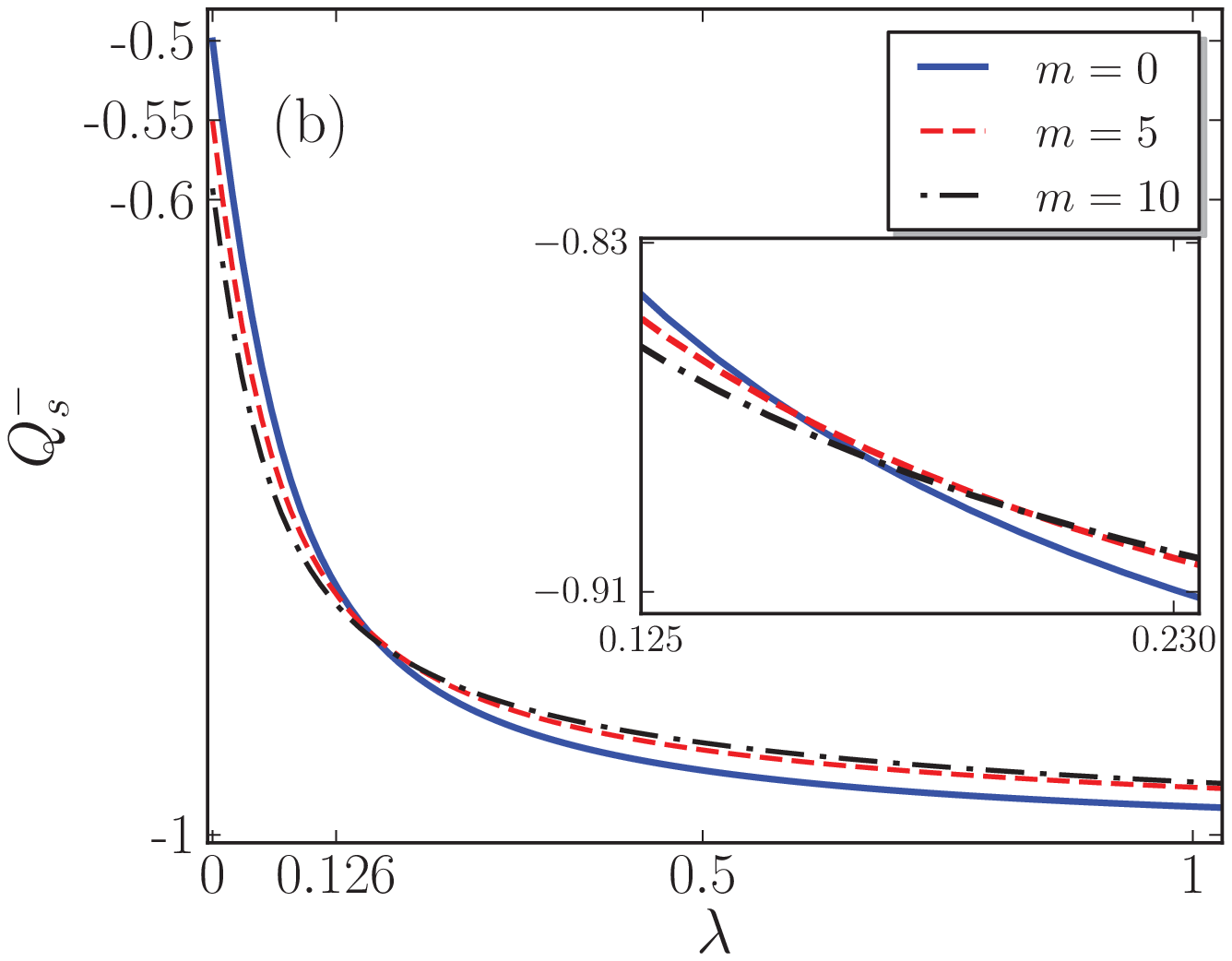}
\end{minipage}
\caption{ Mandel parameter of the state $\vert \mu ,m \rangle^{+}$ (a) and of the state $\vert \mu ,m \rangle^{-}$ (b) versus the curvature $\lambda$ for $N=20$ and $\mu=1$,
 the solid blue curve corresponds to $m=0$, the dashed red curve to $m=5$ and the dot-dashed black curve to $m=10$.}\label{66}
\end{figure*}

In Figure \ref{66}(a), the Mandel parameter is plotted for the PACSs on a sphere in terms of $ \lambda $
for different values of $m$. It is seen that  the Mandel parameter for the state $\vert \mu ,m \rangle^{+} $ decreases by increasing $ {\lambda} $, so that it tends to $-1$ at large curvature.
Moreover, it is observed that for a specific $ {\lambda} $, increasing $m$ reduces the
Mandel parameter. These results can be justified by the fact that the quantum Fock states $ \vert n\rangle $ have the Mandel parameter $ Q=-1 $. Because as seen in Figs. \ref{11}(a) and \ref{11}(b), by increasing $m$ and $ \lambda $, the PACS approaches to the number state $ \vert N\rangle $ which has a Mandel parameter equal to $-1$.

Fig. \ref{66}(b) shows  the Mandel parameter for the PSCSs on a sphere in terms of $ \lambda $. It is seen that the Mandel parameter decreases with increasing $ \lambda $, similar to Fig. \ref{66}(a).
However,  at small (large) values of $\lambda $, an increase
in $m$ leads to decrease (increase) of the Mandel parameter, as shown in inset of Fig. \ref{66}(b).
To explain this behavior, we can recall the effect of increasing $\lambda $ and $m$ in Figs. \ref{44}(a) and \ref{44}(b). The results showed that the increase in $m$ and $ \lambda $ have a reverse effect on $ (P_{n}) _{m}^{-} $.
Hence, as can be seen in Fig. \ref{66}(b), by increasing $ \lambda $, the effect of
$m$ on the Mandel parameter $Q$, is reversed.

\subsection{Quadrature squeezing}

In this section, we are interested to examine the quantum noises of quadrature components of the field compared to the standard coherent states. For this purpose,
$\hat{X} _{1} $ and $\hat{X}_{2}$ are defined in terms of annihilation and creation operators $\hat{a}$  and  $\hat{a}^{\dagger}$ as,
 \begin{eqnarray}\label{X}
    \hat{X}_{1}&=& \frac{1}{2} (\hat{a} e^{i\varphi} + \hat{a}^{\dagger} e^{-i\varphi}),\nonumber\\
    \hat{X}_{2} &=& \frac{1}{2i} (\hat{a} e^{i\varphi} - \hat{a}^{\dagger} e^{-i\varphi}).
 \end{eqnarray}
The commutation relation between $\hat{a}$ and $\hat{a}^{\dagger}$ leads to the following uncertainty relation:
 \begin{equation}
(\triangle {X}_{1})^{2} (\triangle {X}_{2})^{2} \geq \frac{1}{16}\vert \langle [\hat{X}_{1}  ,\hat{X}_{2}]\rangle \vert
^{2}=\frac{1}{16}.
\end{equation}
Now, by definition, the quadrature squeezing occurs in the condition,
 \begin{equation}
    (\triangle \hat{X}_{i})^{2}< \frac{1}{4},\qquad (i=1\ or\ 2),
 \end{equation}
or equivalently
 \begin{equation}\label{S}
    S_{i}\equiv 4(\bigtriangleup X_{i})^{2} -1<0.
 \end{equation}

Now, with the help of Eqs. (\ref{S}) and (\ref{X}), we can investigate the squeezing properties of the PACSs and PSCSs on a sphere. Given the complexity of the resulting equations for the PACSs and PSCSs on a sphere, in the following, we study this nonclassical feature numerically.

\subsubsection{The photon-added}

\begin{figure*}[ht]
\begin{minipage}[b]{0.5\linewidth}
\centering
\includegraphics[width=\textwidth]{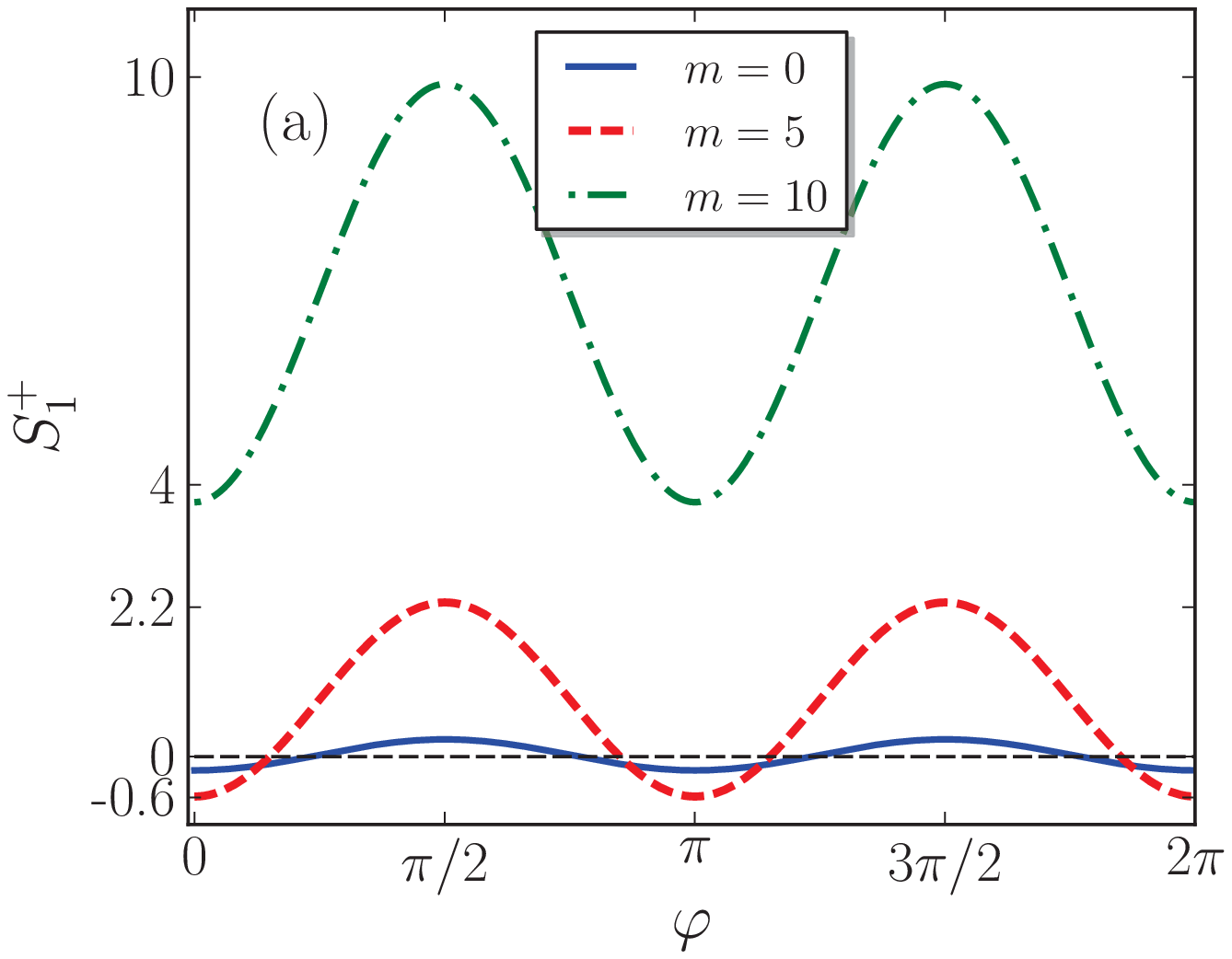}
\end{minipage}
\hspace{0cm}
\begin{minipage}[b]{0.5\linewidth}
\centering
\includegraphics[width=\textwidth]{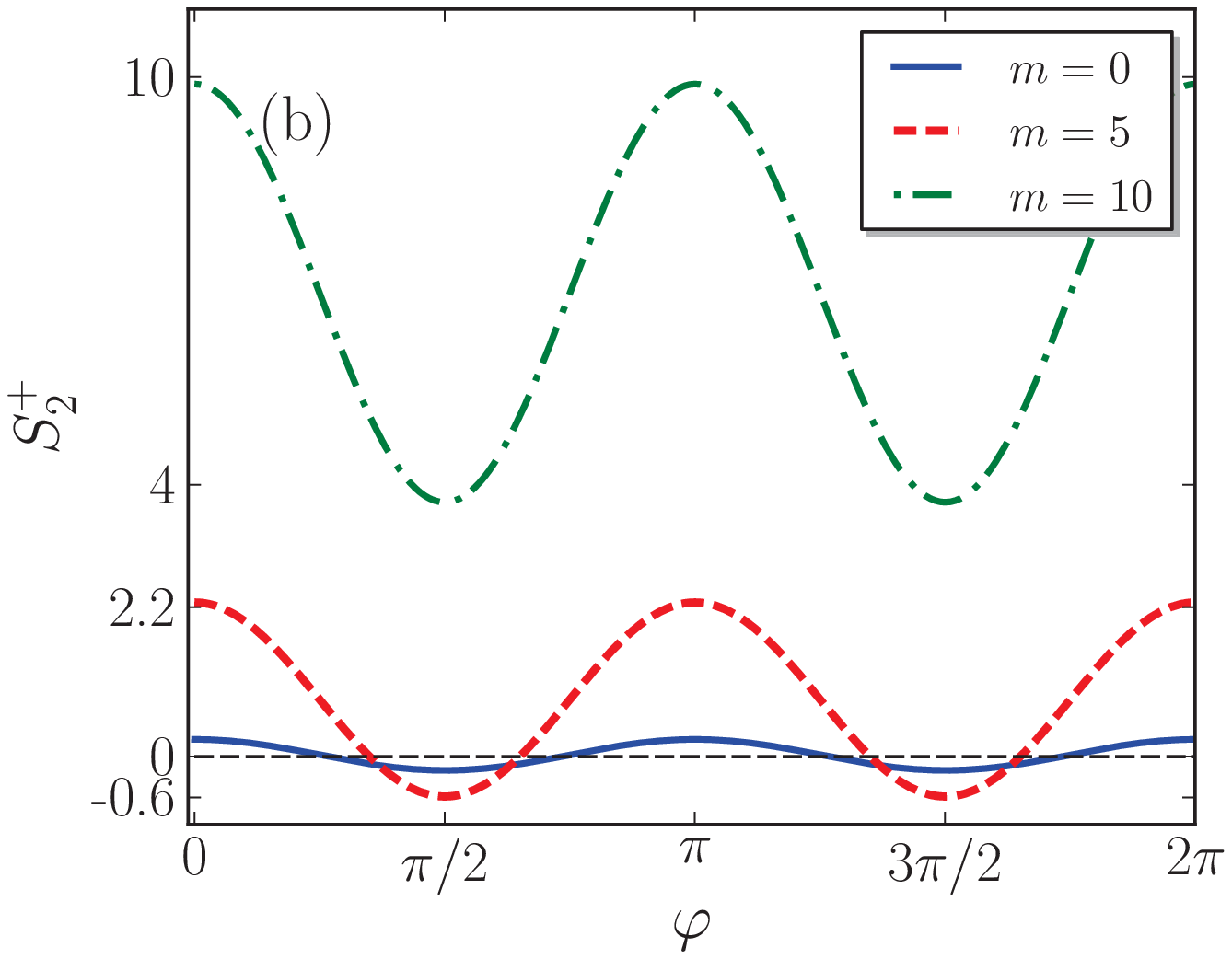}
\end{minipage}
\caption{(a) $S_{1}^{+}$ and (b) $S_{2}^{+}$  for the PACSs on a sphere versus $\varphi$ for $N=20$, $\mu=0.5$ and $\lambda=0$, the solid blue curve corresponds to $m=0$,
the dashed red curve to $m=5$ and the dot-dashed green curve to $m=10$.}\label{88}
\end{figure*}

In Fig. \ref{88}, $ S_{1}^{+}$ and $S_{2}^{+}$ are plotted respectively in terms of $\varphi $ for $N=20$, $\mu =0.5$, $\lambda=0$ and different values of $m$ for the state $\vert \mu ,m\rangle^{+}$.
As can be seen, the degree of squeezing reduces by increasing $m$.

\begin{figure*}[ht]
\begin{minipage}[b]{0.5\linewidth}
\centering
\includegraphics[width=\textwidth]{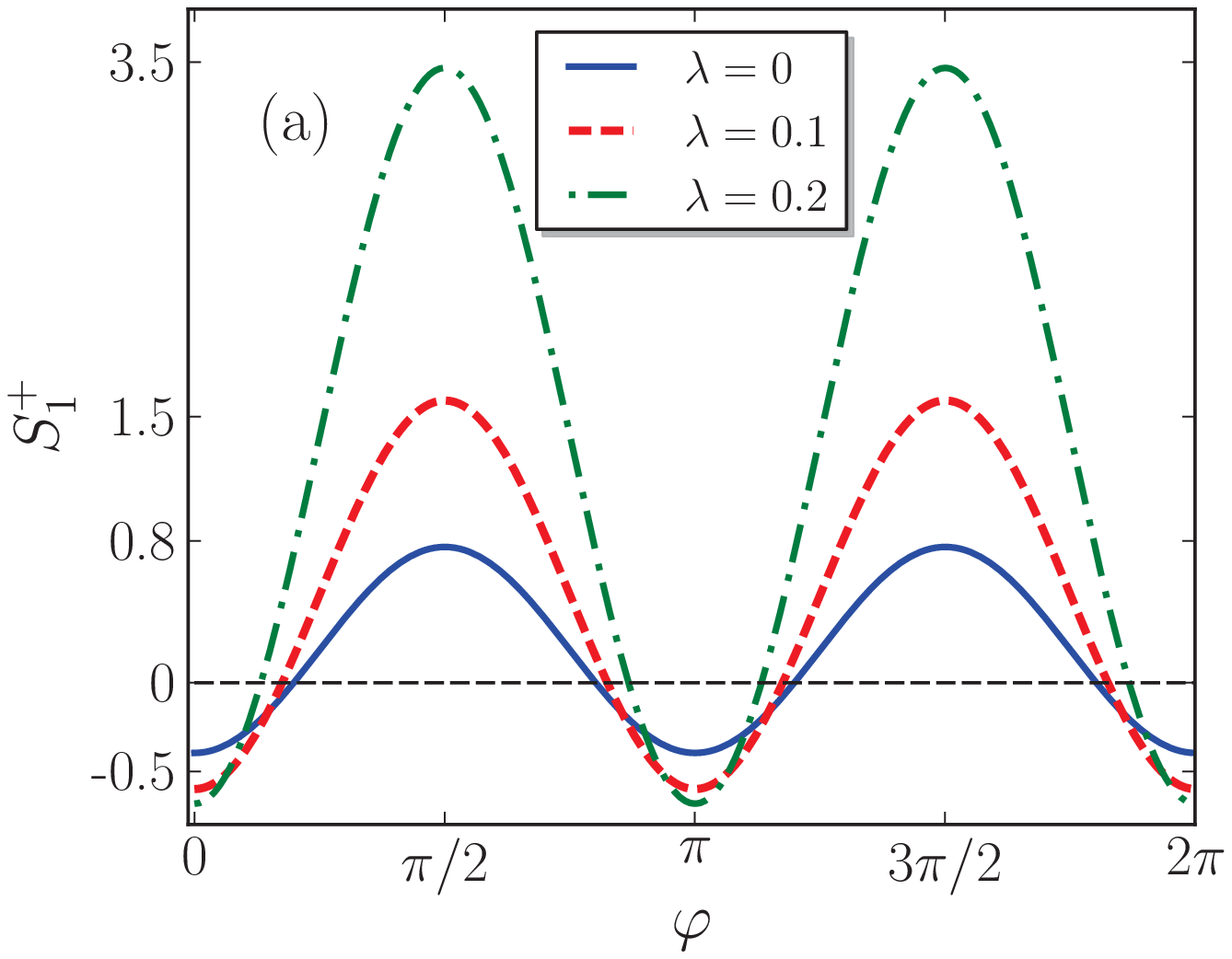}
\end{minipage}
\hspace{0cm}
\begin{minipage}[b]{0.5\linewidth}
\centering
\includegraphics[width=\textwidth]{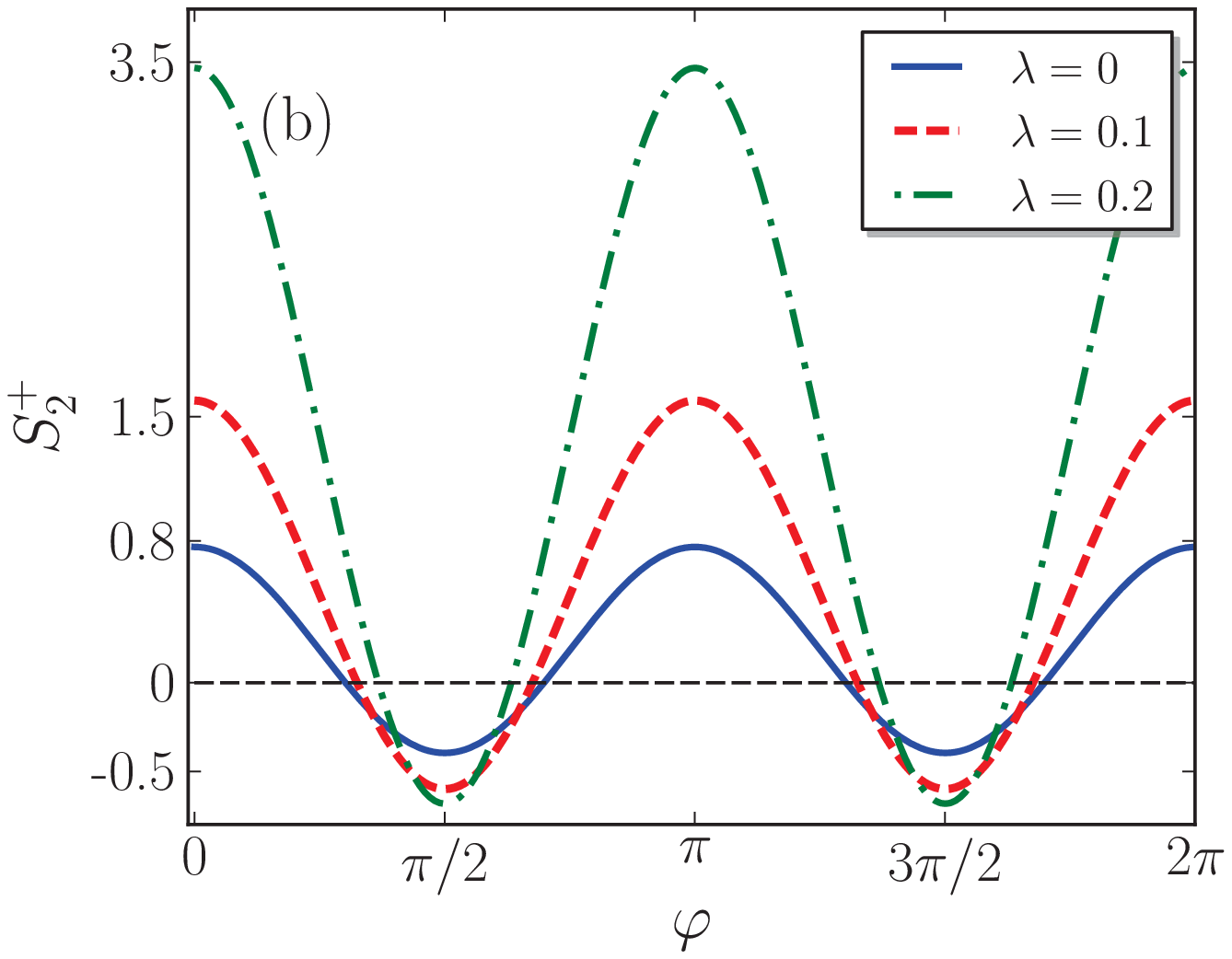}
\end{minipage}
\caption{$S_{1}^{+}$ and $S_{2}^{+}$  for the PACSs on sphere versus $\varphi$ for $N=20$, $\mu=0.5$ and $m=1$, the solid blue curve corresponds to $\lambda =0$, the
dashed red curve to $\lambda =0.1$ and the dot-dashed green curve to $\lambda =0.2$.}\label{99}
\end{figure*}

Fig. (\ref{99}), respectively, show $ S_{1}^{+}$ and $S_{2}^{+}$ for $N=20$, $\mu =0.5$, $m =1$ and various values of $ \lambda$.
These figures clearly show that the degree of squeezing is enhanced by increasing $\lambda $.

\subsubsection{The photon-subtracted}

\begin{figure*}[ht]
\begin{minipage}[b]{0.5\linewidth}
\centering
\includegraphics[width=\textwidth]{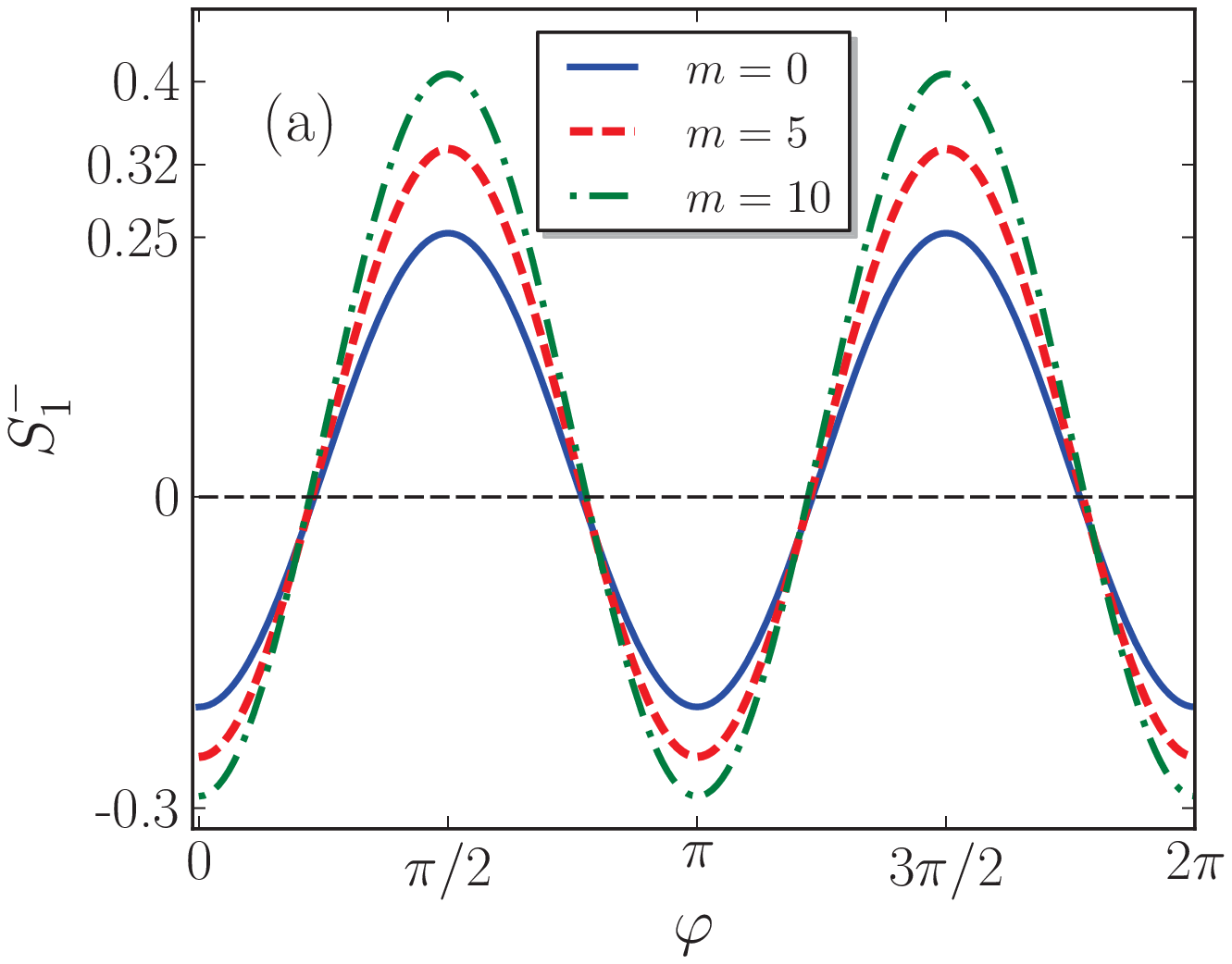}
\end{minipage}
\hspace{0cm}
\begin{minipage}[b]{0.5\linewidth}
\centering
\includegraphics[width=\textwidth]{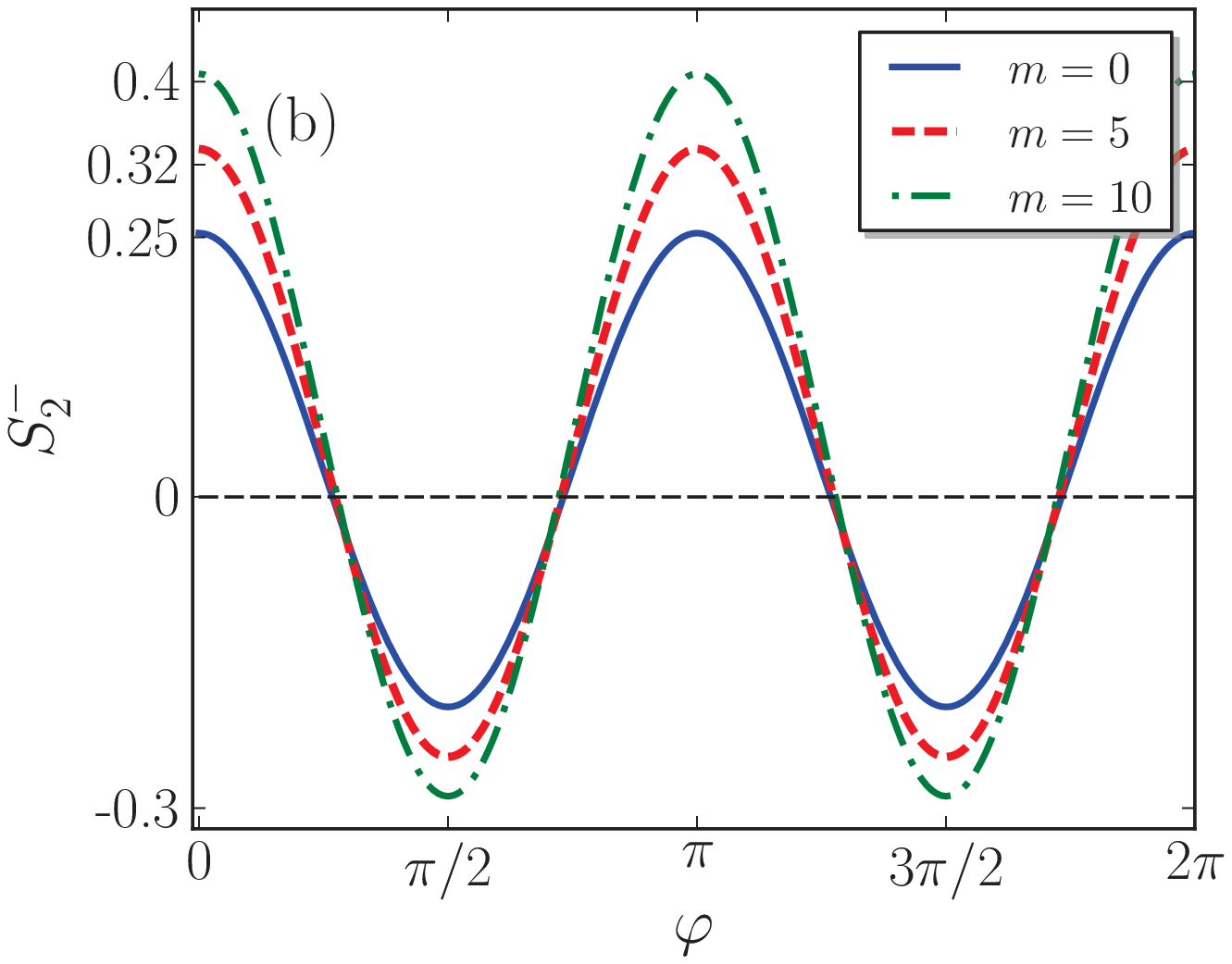}
\end{minipage}
\caption{(a) $S_{1}^{-}$ and $(b)$ $S_{2}^{-}$  for the PSCSs on sphere versus $\varphi$ for $N=20$, $\mu=0.5$ and $\lambda=0$, the solid blue curve corresponds to $m=0$,
the dashed red curve to $m=5$ and the dot-dashed green curve to $m=10$.}\label{1010}
\end{figure*}

In Fig. (\ref{1010}), $ S_{1}^{-}$ and $S_{2}^{-}$ are plotted respectively in terms of $\varphi $  for $N=20$, $\mu =0.5$, $\lambda=0$ and different values of $m$ for  the
PSCSs on a sphere.
Here, we find that by increasing $m$, the degree of squeezing is enhanced.

\begin{figure*}[ht]
\begin{minipage}[b]{0.5\linewidth}
\centering
\includegraphics[width=\textwidth]{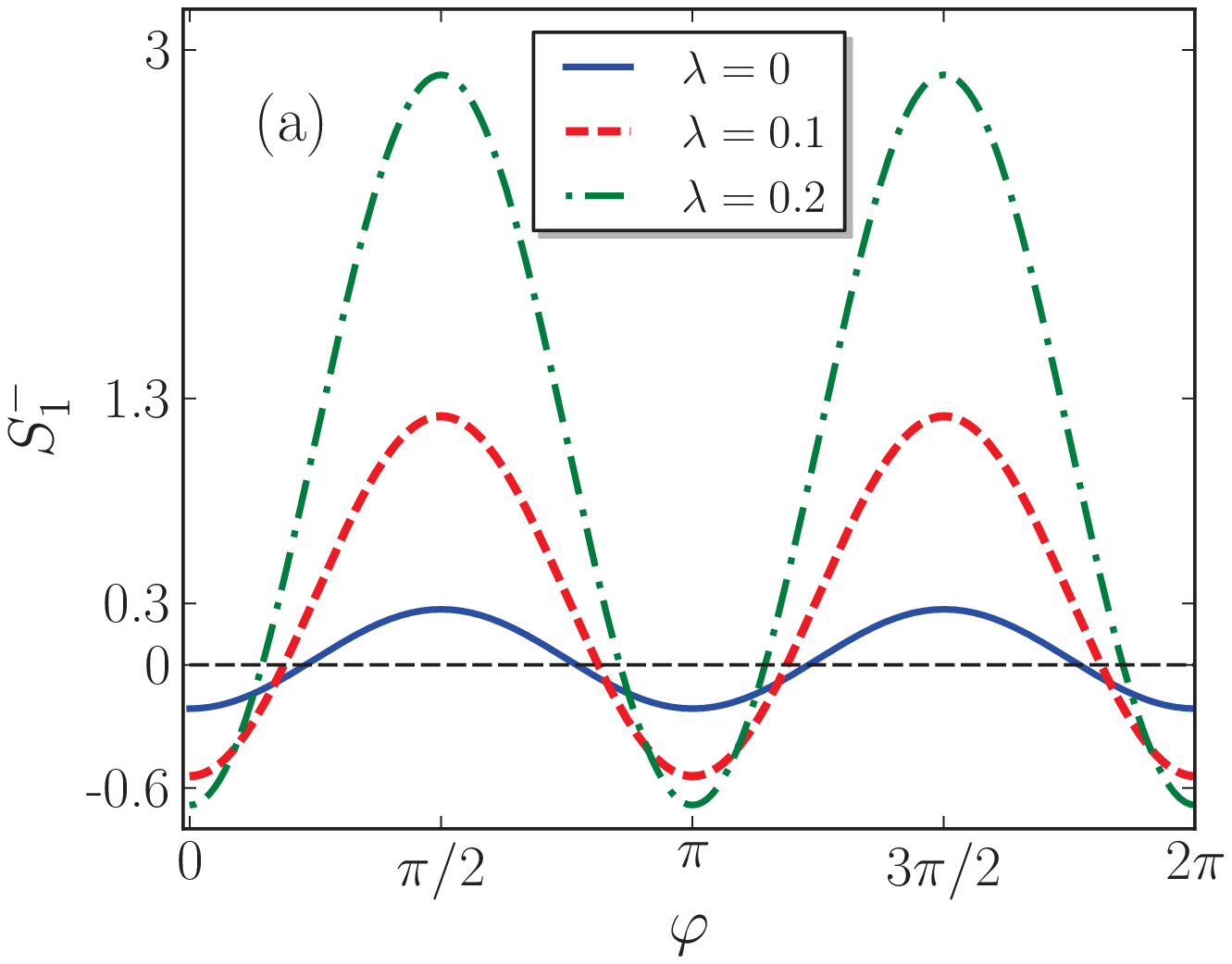}
\end{minipage}
\hspace{0cm}
\begin{minipage}[b]{0.5\linewidth}
\centering
\includegraphics[width=\textwidth]{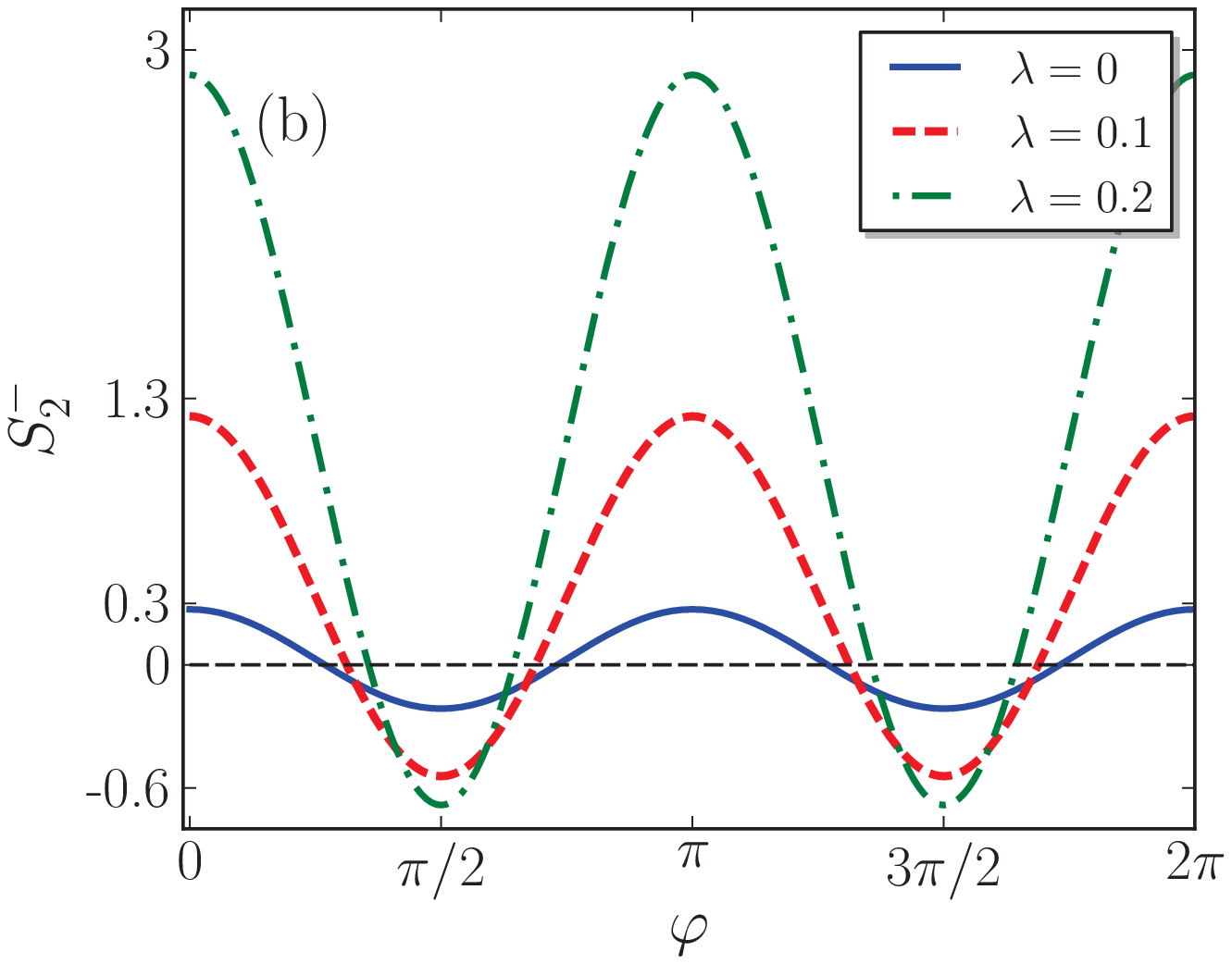}
\end{minipage}
\caption{$S_{1}^{-}$ and $S_{2}^{-}$  for the PSCSs on sphere versus $\varphi$ for $N=20$, $\mu=0.5$ and $m=1$, the solid blue curve corresponds to $\lambda =0$, the
dashed red curve to $\lambda =0.1$ and the dot-dashed green curve to $\lambda =0.2$.}\label{1111}
\end{figure*}

Fig. (\ref{1111}), respectively, display $ S_{1}^{-}$ and $S_{2}^{-}$ in terms of $\varphi$ for $N=20$, $\mu =0.5$, $m=1$ and different values of $\lambda$.
These figures obviously show that by increasing $\lambda$, the quadrature squeezing is enhanced.

\section{Summary and concluding remarks}\label{Sum}
In this paper, we have constructed the $m$-PACSs and $m$-PSCSs on a sphere and proved that these states form an overcomplete set. We have also presented a physical scheme to generate PACSs and PSCSs on a sphere, as the states of radiation field, for different values of the $m$ and the curvature.
Furthermore, we have considered the non-classical behaviors of the aforementioned states, by using the photon mean number, the Mandel parameter and the quadrature squeezing. The results
show that the PACSs and PSCSs on a sphere have sub-Poissonian distributions and the degree of squeezing is reduced (enhanced) by increasing $m$ for the photon-added (photon-subtracted) coherent states on a sphere.
The results also exhibit that the curvature of physical space leads to the enhancement of nonclassical properties of the states.

%//////////////////////////////////////////////////////////////////////////////////////////

\end{document}